\documentstyle[12pt]{article}
\title{The LIGO Research Community: Defining a Research Environment}
\author{Lee Samuel Finn\\
Chairperson, LIGO Research Community\\
\tt lrc@ligo.caltech.edu}
\date{17 January 1996}
\begin{document}
\maketitle
\section{Introduction}

The LIGO Research Community (LRC) held its first meeting at the Aspen
Center for Physics on 17 January 1996. At the meeting we initiated a
several month study to identify how LIGO's completion will affect the
research environment in gravitational physics and recommend, to LIGO
and the NSF, policies to be adopted now (on, {\em e.g.,} data
availability, proposals for new or enhanced interferometer, scheduling
of interferometer time, research internships at LIGO facilities, {\em
etc.}) in order to make the most of the opportunities presented.

This report describes the LRC --- its aims and goals, membership and
officers, and how to join --- and also the LRC study program
``Defining A Research Environment: LIGO and the Gravitational Physics
Research Community'' --- what it involves and how to participate. I
urge everyone with an interest in gravitational wave research to read
especially \S\ref{sec:defining} and contribute to the study their
thoughts and concerns regarding the impact that LIGO will have on the
future shape of the gravitational physics research environment.

\section{What is the LIGO Research Community?}
\subsection{Aims and Goals}

The LRC Charter identifies the principal aims and goals of the LIGO
Research Community:
\begin{itemize}
\item To provide an organized channel for the interchange of
information between the LIGO management and those individuals who
utilize the scientific opportunities afforded by LIGO; and
\item To serve as an advocacy body for the study of gravitational
waves and related physics and astronomy.
\end{itemize}

Within the research community, the LRC exists to facilitate exchange
between those who build the hardware to detect, those who develop the
data analysis tools and techniques to recognize, and those who explore
the theory of gravitational waves and their sources. In its
relationship with LIGO, the LRC's role is to allow two-way
communication between the LIGO management and the broader research
community about design decisions now and scientific program decisions
in the future, to insure that LIGO functions as part of an {\em
international} gravitational research community, and to nurture a body
of people who care enough about LIGO that they would insist on it
living up to its goals and promises, and protest vigorously if it
doesn't.

In short, the LRC is the voice of the research community with a
professional or personal interest in LIGO operations or related
science.

\subsection{What is the relationship of the LRC to LIGO?}

The LRC is an independent organization of individuals with a common
interest in the science, technology or operation of LIGO. It is not an
arm of the LIGO project and is not affiliated with LIGO, though it
gratefully acknowledges the administrative support of the LIGO
project.

\subsection{Members}

At the close of the election of the first LRC executive committee
there were  175 members of the LIGO Research Community representing
19 countries (see table \ref{tbl:demographics} for details).
\begin{table}
\caption{LIGO Research Community Demographics as of 6 November. Of the
95 US members, 12 are part of the LIGO research
effort.}\label{tbl:demographics}
\begin{tabular}{||l|r||l|r||}
\hline\hline
Country&Number&Country&Number\\
\hline
Australia&1&Brazil&3\\
Canada&4&Denmark&1\\
Germany&13&Greece&1\\
India&6&Ireland&1\\
Italy&7&Japan&14\\
Mexico&2&Netherlands&4\\
Poland&2&Russia&3\\
South Africa&2&Spain&1\\
Taiwan&1&United Kingdom&14\\
United States&95&&\\
\hline\hline
\end{tabular}
\end{table}

\subsection{Officers}

The LRC is governed by an executive committee, composed of seven
members elected by and from its membership for three-year terms. The
executive committee elects from its voting members the LRC Chair. The
95-96 executive committee consists of
\begin{itemize}
\item Bruce Allen (University of Wisconsin, Milwaukee; early universe
cosmology, gravitational wave data analysis),
\item Joan Centrella (Drexel University; sources of gravitational
radiation, numerical relativity and astrophysics)
\item Sam Finn\footnote{LIGO Research Community Chair} (Northwestern
University; general relativistic astrophysics, gravitational wave data
analysis, gravitational wave astronomy)
\item Eric Gustafson (Stanford University; laser systems, experimental
gravitational wave detection)
\item Bill Hamilton (Louisiana State University, Baton Rouge;
experimental gravitational wave detection)
\item David Shoemaker (Massachusetts Institute of Technology and LIGO
Project; experimental gravitational wave detection)
\item Harry Ward (University of Glasgow and GEO 600 Project;
experimental gravitational wave detection).
\end{itemize}
In addition to the seven elected members of the executive committee,
the LIGO principal investigator is a non-voting, {\em ex officio}
member of the executive committee, and the {\em Executive Committee
Secretary} (currently Syd Meshkov) is appointed by the chairperson to
aid the committee's work.

To contact the Executive Committee with questions or comments, send
e-mail to {\tt lrc\_excomm@holmes.astro.nwu.edu}.

\subsection{How to Join}

The LIGO Research Community is an organization of individuals and
membership is open to any individual, irrespective of other
affiliations. To join the LRC, send an e-mail message to {\tt
lrc@ligo.caltech.edu} indicating your desire to become a member.

Membership entitles you to vote for, run for and hold LRC
offices. Membership must be renewed yearly via an e-mail note to
{\tt lrc@ligo.caltech.edu}. You are entitled to vote in an LRC election if
you have joined or renewed your membership between the end of the last
election and the close of nominations for the current election.

Since the LRC is an organization of individuals, there are no
institutional members. The LRC may form liaisons with other groups
pursuing common goals, and members of other groups may become members
of the LRC; however, groups cannot themselves become LRC members.

\section{Defining a Research Environment: LIGO and the Gravitational
Physics Research Community\label{sec:defining}}

\subsection{Introduction}

The construction of LIGO and VIRGO --- gravitational wave detectors
whose ultimate sensitivity virtually insures a steady rate of
detections --- has had a profound effect on the research environment
in gravitational physics and astrophysics, creating new research
opportunities and giving new relevance to old lines of research.  Many
of us --- both ``theoretical'' and ``experimental'' physicists ---
have redirected our research programs to take advantage of these new
opportunities. As the detectors come on-line --- as data begins to
flow and as interferometer enhancements and entirely new
interferometers are developed --- our research environment will evolve
even more extensively. What will be the nature of the research environment
that we will find ourselves working in five or ten years from now?
\begin{itemize}
\item How, and to who, does an experimental group propose to develop
new instrumentation for use in the LIGO facility? How does a group
gain the necessary expertise to develop a new instrument, or a
modification of an existing one? What infrastructure can be counted
on, and what facilities must be part of the proposal (e.g., who owns
the laser light? who owns the vacuum chambers? who owns the mirrors
and test masses?)
\item How does a theorist gain access to data, or contribute to data
analysis? If LIGO is counting on the non-LIGO research community to
develop and contribute data analysis tools, how are those tools
vetted?
\item How are the operating priorities of the several LIGO
interferometers set: {\em i.e.,} what fraction of the operating time
is devoted to engineering {\em vs.} two or three interferometer
coincidence ``science runs?''
\item What steps must be taken to insure that LIGO operates as a
constructive part of an international community of gravitational wave
detectors ({\em e.g.,} common data formats and standards, cooperative
agreements for data exchange with VIRGO, {\em etc.})?
\item What support does the NSF need to provide the research community
so that the most is made of the opportunities LIGO provides?
\end{itemize}

There are many possible futures, but the research environment that we
find ourselves in five or ten years from now will arise from the
actions we take and the decisions we make today. Now we have the
opportunity to identify the kind of research environment we want and
work to make it happen. To that end, the LRC has begun a several month
study to
\begin{itemize}
\item {\em Identify} how an operating LIGO will affect our research
environment,
\item {\em Decide} what we want the research environment to look like,
and
\item {\em Recommend,} to LIGO and the NSF, what steps should be taken
to create that environment
\end{itemize}
The first part of this program was initiated at Aspen with the
formation of three {\em Issue Identification Groups,} which are
described in more detail below. Through the end of March these groups
will canvas the research community with the goal of identifying how
our research environment will be transformed by the initiation of
LIGO/VIRGO operations\footnote{One need not be a member of the LRC to
contribute to this study.}. By the middle of April these groups will
report to the executive committee, identifying and discussing those
issues highlighted by the research community.

In response to the IIG reports, the executive committee will create
several policy study committees to address the most pressing
issues. These committees will be announced at the next general
membership meeting of the LRC, which will take place at the May
meeting of the American Physical Society (Indianapolis, Indiana, 2-5
May 1996). Each committee will be charged with canvassing the
research community and reporting back to the executive committee with
recommendations that reflect the combined judgment of the committee
members and the research community. Using these reports as a basis,
the executive committee will prepare a report for the National Science
Foundation and the LIGO project that recommends policies to be adopted
now --- by ourselves, the LIGO project, and the National Science
Foundation --- to insure that the best is made of the new
opportunities that await us.

The three Issue Identification Groups are
\begin{itemize}
\item Sources, data and analysis,
\item Hardware development, installation and operations, and
\item People.
\end{itemize}
The focus of each group is discussed in more detail below. The group
chairs are responsible for selecting a small set of members from
research community volunteers to assist in canvassing the membership
and preparing the list of issues.

Contributions to each group are critical to the success of this
program. One need not be a member of the LRC in order to contribute to
this study, and we hope that all who are interested or involved in
gravitational physics research will contribute their thoughts on how
the operation of the LIGO and VIRGO detectors will change the
gravitational physics research environment.  All contributions will
appear in the final IIG reports. Each IIG has a dedicated e-mail
address (provided below) where you are encouraged to send
contributions or volunteer as ;a working group member.

\subsection{Sources, Data and Analysis}
\begin{itemize}
\item Chair: Joan Centrella.
\item E-Mail: {\tt lrc\_sda@holmes.astro.nwu.edu}
\end{itemize}
The focus of this group is data issues, data analysis issues, or
source calculation issues. Example issues include
\begin{itemize}
\item Data access: should data be made generally available
immediately, never, after a period of proprietary access by the
experimental team responsible for the instruments, or by proposal?
What are the consequences of each possibility?
\item Computational needs: Can the computational needs of the data
analysis task be met without dedicated supercomputer facilities? If
not, how are these facilities provided for?
\item Cooperation with other facilities: How tightly should LIGO's data
format and policies be coupled with those of VIRGO and other detector
facilities?
\end{itemize}

\subsection{Hardware Development, Installation and Operations}
\begin{itemize}
\item Chair: Eric Gustafson.
\item E-Mail: {\tt lrc\_hardware@holmes.astro.nwu.edu}
\end{itemize}
The focus of this group is issues related to the operation of
installed interferometers and the development of enhancements and
advanced instrumentation. Example issues include
\begin{itemize}
\item Hardware proposals: how does a research group propose an
instrumentation enhancement or a new interferometer,{\em i.e.,} who
does one propose to? how is a proposal reviewed? how is a proposal
funded? what must a proposal contain? what infrastructure is provided
by LIGO, and what must be included in a proposal?
\item Operations: how is interferometer time scheduled, {\em i.e.,}
how are the demands of engineering balanced against double and triple
coincidence ``science'' runs?
\item How does a non-LIGO experimental team develop the expertise
necessary to develop an interferometer enhancement or propose a new
detector?
\end{itemize}

\subsection{People}
\begin{itemize}
\item Chair: Harry Ward
\item E-Mail: {\tt lrc\_people@holmes.astro.nwu.edu}
\end{itemize}
The focus of this group is both ``human resource'' and
inter-organizational issues. Examples include
\begin{itemize}
\item Visitors Program: how are long-term visits to LIGO facilities,
{\em e.g.,} for the purpose of acquiring expertise to take back to
one's home institution, proposed for/funded?
\item LIGO staff liaisons: how do LIGO ``data product users''
correspond with instrument teams to understand the data stream? How do
experimenters at other institutions work with LIGO staff in order that
their instruments integrate properly in the LIGO facilities?
\item What should be the relationship between the LRC and other
organizations? Between the LIGO and other detector projects?
\end{itemize}

\section{Conclusion}

The LIGO Research Community (LRC) is an independent organization of
researchers interested in the scientific opportunities created by the
construction and operation of the Laser Interferometer
Gravitational-wave Observatory (LIGO). Membership is open to all
interested individuals, irrespective of any other affiliations
(including affiliation with the LIGO project, VIRGO or other
gravitational-wave detector projects). The LRC is engaged in a study
project designed to {\em identify} the ways that an operating LIGO
will affect the research environment in gravitational physics, {\em
decide} what we want that environment to look like, and recommend (to
LIGO and the NSF) the steps to be taken now to develop that
environment in the future. Contributions from LRC members and from the
broader gravitational physics research community are actively
solicited.

\end{document}